\documentclass[conference]{IEEEtran}
\IEEEoverridecommandlockouts
\def\review{0} %
\if\review1
\usepackage{todonotes}
\else
\usepackage[disable]{todonotes}
\fi
\usepackage{cite}
\usepackage{amsmath,amssymb,amsfonts}
\usepackage{graphicx}
\usepackage{textcomp}
\usepackage{xcolor}
\usepackage{makecell}
\usepackage{multicol}
\usepackage{siunitx}
\usepackage{booktabs}
\usepackage{xfrac}
\usepackage{algorithm}
\usepackage{algpseudocode}
\usepackage[nolist]{acronym}
\usepackage{tikz}
\usepackage{pgfplots}
\usepackage{rotating}
\pgfplotsset{width=7cm,compat=1.3}
\usetikzlibrary{shapes,arrows,positioning,calc, matrix,spy,patterns}
\usetikzlibrary{decorations.pathreplacing,calligraphy}
\usetikzlibrary{colorbrewer}
\usetikzlibrary{fit}
\usepgfplotslibrary{colorbrewer}
\usepackage{bbm}
\usepackage{bm}
\usepackage{mathtools}
\usepackage{pgffor}
\usetikzlibrary{fit,backgrounds}
\usepgfplotslibrary{fillbetween}
\usepackage{dblfloatfix}
\usepackage{url}

\definecolor{cb-1}{HTML}{4477AA}
\definecolor{cb-2}{HTML}{EE6677}
\definecolor{cb-3}{HTML}{228833}
\definecolor{cb-4}{HTML}{CCBB44}
\definecolor{cb-5}{HTML}{66CCEE}
\definecolor{cb-6}{HTML}{AA3377}
\definecolor{cb-7}{HTML}{BBBBBB}

\pgfplotscreateplotcyclelist{cb list}{
	cb-1,cb-2,cb-3,cb-4,cb-5,cb-6,cb-7
}

\definecolor{kit-green100}{rgb}{0,.59,.51}
\definecolor{kit-green70}{rgb}{.3,.71,.65}
\definecolor{kit-green50}{rgb}{.50,.79,.75}
\definecolor{kit-green30}{rgb}{.69,.87,.85}
\definecolor{kit-green15}{rgb}{.85,.93,.93}
\definecolor{KITgreen}{rgb}{0,.59,.51}

\definecolor{KITpalegreen}{RGB}{130,190,60}
\colorlet{kit-maigreen100}{KITpalegreen}
\colorlet{kit-maigreen70}{KITpalegreen!70}
\colorlet{kit-maigreen50}{KITpalegreen!50}
\colorlet{kit-maigreen30}{KITpalegreen!30}
\colorlet{kit-maigreen15}{KITpalegreen!15}

\definecolor{KITblue}{rgb}{.27,.39,.66}
\definecolor{kit-blue100}{rgb}{.27,.39,.67}
\definecolor{kit-blue70}{rgb}{.49,.57,.76}
\definecolor{kit-blue50}{rgb}{.64,.69,.83}
\definecolor{kit-blue30}{rgb}{.78,.82,.9}
\definecolor{kit-blue15}{rgb}{.89,.91,.95}

\definecolor{KITyellow}{rgb}{.98,.89,0}
\definecolor{kit-yellow100}{cmyk}{0,.05,1,0}
\definecolor{kit-yellow70}{cmyk}{0,.035,.7,0}
\definecolor{kit-yellow50}{cmyk}{0,.025,.5,0}
\definecolor{kit-yellow30}{cmyk}{0,.015,.3,0}
\definecolor{kit-yellow15}{cmyk}{0,.0075,.15,0}

\definecolor{KITorange}{rgb}{.87,.60,.10}
\definecolor{kit-orange100}{cmyk}{0,.45,1,0}
\definecolor{kit-orange70}{cmyk}{0,.315,.7,0}
\definecolor{kit-orange50}{cmyk}{0,.225,.5,0}
\definecolor{kit-orange30}{cmyk}{0,.135,.3,0}
\definecolor{kit-orange15}{cmyk}{0,.0675,.15,0}

\definecolor{KITred}{rgb}{.63,.13,.13}
\definecolor{kit-red100}{cmyk}{.25,1,1,0}
\definecolor{kit-red70}{cmyk}{.175,.7,.7,0}
\definecolor{kit-red50}{cmyk}{.125,.5,.5,0}
\definecolor{kit-red30}{cmyk}{.075,.3,.3,0}
\definecolor{kit-red15}{cmyk}{.0375,.15,.15,0}

\definecolor{KITpurple}{RGB}{160,0,120}
\colorlet{kit-purple100}{KITpurple}
\colorlet{kit-purple70}{KITpurple!70}
\colorlet{kit-purple50}{KITpurple!50}
\colorlet{kit-purple30}{KITpurple!30}
\colorlet{kit-purple15}{KITpurple!15}

\definecolor{KITcyanblue}{RGB}{80,170,230}
\colorlet{kit-cyanblue100}{KITcyanblue}
\colorlet{kit-cyanblue70}{KITcyanblue!70}
\colorlet{kit-cyanblue50}{KITcyanblue!50}
\colorlet{kit-cyanblue30}{KITcyanblue!30}
\colorlet{kit-cyanblue15}{KITcyanblue!15}

\definecolor{KITbraun}{RGB}{167,130,46}

\definecolor{cb-1}{HTML}{4477AA}
\definecolor{cb-2}{HTML}{EE6677}
\definecolor{cb-3}{HTML}{228833}
\definecolor{cb-4}{HTML}{CCBB44}
\definecolor{cb-5}{HTML}{66CCEE}
\definecolor{cb-6}{HTML}{AA3377}
\definecolor{cb-7}{HTML}{BBBBBB}

\pgfplotscreateplotcyclelist{cb list}{
	cb-1,cb-2,cb-3,cb-4,cb-5,cb-6,cb-7
}

\newcommand{\RV}[1]{\mathsf{#1}}

\newcommand*{\e}{\mathrm{e}}
\let\j\relax
\newcommand{\j}{\mathrm{j}}

\usepackage{xcolor}
\usepackage{amsmath}
\usepackage{amssymb}
\usetikzlibrary{calc}
\usetikzlibrary{positioning}
\usetikzlibrary{patterns}
\usetikzlibrary{decorations.pathreplacing,calligraphy}

\tikzset{
	lsblock/.style={rectangle, thick, draw, minimum width=1.2cm, minimum height=0.6cm, rounded corners=1.6mm, font=\small, align=center}
}

\tikzset{
	aeblock/.style={rectangle, thick, draw, minimum width=2.4cm, minimum height=0.7cm, rounded corners=1.6mm, align=center}
}
\tikzset{
	smlblock/.style={rectangle, thick, draw, minimum width=1.8cm, minimum height=0.6cm, rounded corners=1.2mm, align=center}
}
\tikzset{
	smsblock/.style={rectangle, thick, draw, minimum width=0.6cm, minimum height=0.4cm, rounded corners=1.0mm,inner sep=0.05cm, align=center}
}

\tikzstyle{surround} = [rectangle, rounded corners, draw=KITred, inner sep=0.25cm, dashed, thick]
\tikzstyle{surround_AE} = [rectangle, rounded corners, draw=KITred, inner sep=0.13cm, dashed, thick]

\def\BibTeX{{\rm B\kern-.05em{\sc i\kern-.025em b}\kern-.08em
    T\kern-.1667em\lower.7ex\hbox{E}\kern-.125emX}}

\begin{document}
\begin{acronym}[]
    \acro{6G}{sixth generation}
    \acro{AE}{autoencoder}
    \acro{AESL}{average expected sidelobe level}
    \acro{AIR}{achievable information rate}
    \acro{AWGN}{additive white Gaussian noise}
    \acro{CP}{cyclic prefix}
    \acro{DFT}{discrete Fourier transform}
    \acro{MI}{mutual information}
    \acro{IDFT}{inverse discrete Fourier transform}
    \acro{ISAC}{integrated sensing and communications}
    \acro{LLR}{log-likelihood ratio}
    \acro{MF}{matched filter}
    \acro{NN}{neural network}
    \acro{OFDM}{orthogonal frequency division multiplexing}
    \acro{PA}{power allocation}
    \acro{PDF}{probability density function}
    \acro{PMF}{probability mass function}
    \acro{PSK}{phase shift keying}
    \acro{QAM}{quadrature amplitude modulation}
    \acro{RV}{random variable}
    \acro{SC}[S\&C]{sensing \& communications}
    \acro{SI}{self interference}
    \acro{SL}{sidelobe level}
    \acro{ISINR}{image signal-to-interference-and-noise ratio}
    \acro{SNR}{signal-to-noise ratio}
    \acro{TOI}{target of interest}
    \acro{WF}{waterfilling}
\end{acronym}

\bstctlcite{IEEEexample:BSTcontrol}

\title{Joint Power and Adaptive-Constellation \hspace*{0.6em}Optimization for OFDM-ISAC Systems
\thanks{\IEEEauthorrefmark{1}Both authors contributed equally to this work.}
\thanks{This work has received funding from the German Federal Ministry of Research, Technology and Space (BMFTR) within the project Open6GHub+ (grant agreement 16KIS2405).}
}

\author{
\IEEEauthorblockN{Benedikt Geiger\IEEEauthorrefmark{1}, Linus Hampel\IEEEauthorrefmark{1}, Laurent Schmalen}
\IEEEauthorblockA{Communications Engineering Lab (CEL), Karlsruhe Institute of Technology (KIT), Hertzstr. 16, 76187 Karlsruhe, Germany\\
Email: \texttt{benedikt.geiger@kit.edu}}
}

\maketitle

\begin{abstract}
Next-generation wireless systems are expected to support \ac{ISAC}. However, sensing imposes different requirements on the \ac{PA} and the modulation format than communications, leading to a fundamental trade-off in frequency-selective channels. In this paper, we jointly optimize the \ac{PA} and modulation format for \ac{OFDM}-based \ac{ISAC} systems, considering both target detection and tracking. We first derive an upper bound on the mutual information to predict the communication performance for power-constellation allocation. We then use an \ac{AE} to learn a family of finite constellations that realizes the allocation under different sensing constraints without retraining. The results show that tracking alleviates the \ac{SC} trade-off compared with initial detection, and that the \ac{PA} and the modulation dominate the \ac{SC} trade-off at low and high \acp{SNR}, respectively. At intermediate \ac{SNR}, the proposed joint optimization significantly outperforms individual optimizations, while the pretrained adaptive constellation family almost fully realizes the predicted allocation gains.
\end{abstract}


\acresetall
\vspace{-0.1cm}
\section{Introduction}
\vspace{-0.05cm}
In future wireless systems, \ac{ISAC} is expected to enable the reuse of the \ac{OFDM} communication signal for radar-like sensing~\cite{liu_integrated_2022}. However, the optimal \ac{PA} and modulation format for \ac{SC} differ~\cite{xiong_torch_2024}. For communications, a \ac{WF} \ac{PA} in combination with a Gaussian-distributed input maximizes the \ac{MI} over an \ac{AWGN} channel under a transmit power constraint~\cite{Cover2006}. In contrast, a uniform \ac{PA} and unit-modulus modulation format reduce ranging \acp{SL} of the autocorrelation function~\cite{zhang_optimal_2025}.

Therefore, constellation shaping has recently gained attention as a means to balance \ac{SC} performance~\cite{du_reshaping_2023,Geiger25JCS}. However, these works optimize constellations for fixed sensing constraints, leaving adaptive constellation shaping for varying \ac{SC} requirements unexplored. Furthermore, the impact of the \ac{PA} on sensing has been characterized and optimized for initial target detection, where the \acp{SL} averaged over the delay domain are minimized~\cite{zhang_optimal_2025,yilmaz_rate_exponent_2026}. However, it remains unclear to what extent prior target information can alleviate the \ac{SC} trade-off by localizing the sensing constraint to a small delay region.

While the individual modulation- and power-induced trade-offs have been studied extensively, the literature on their joint optimization remains sparse. To the best of our knowledge, existing joint approaches are limited to a power and constellation selection framework~\cite{meng_constellation_2026} and an information-theoretic continuous input distribution design based on an upper bound~\cite{zhao_input_2026}. These works provide important initial insights into the coupled power-modulation trade-off. However, selecting constellations from a finite set of conventional modulation formats underestimates the achievable \ac{SC} performance, whereas continuous input distribution design provides an upper benchmark that overestimates the performance attainable with practical finite constellations. Moreover, both approaches mainly focus on the high \ac{SNR} regime, where the \ac{PA}-induced trade-off largely vanishes, leaving the coupled power-modulation trade-off insufficiently characterized.

In this paper, we bridge the gap between conventional modulation format selection and continuous input distribution design. Specifically, we jointly optimize the \ac{PA} and modulation for \ac{OFDM}-\ac{ISAC} systems operating in frequency-selective channels. To make the joint optimization tractable, we generalize the upper bound on the \ac{MI} from~\cite{geiger_constellation_2025} to account for frequency-selective fading and \ac{PA}. Since both the resulting \ac{MI} surrogate and the sensing metric depend only on power and kurtosis, i.e., the fourth-order moment of the constellation, we split the original problem into constellation shaping and power-kurtosis allocation. We then use an \ac{AE} framework to optimize offline a parameterized family of finite constellations, enabling \ac{MI}-optimized constellations to be generated online for the allocated kurtosis values without retraining. Finally, simulation results validate the proposed approach: The joint optimization outperforms optimizations over either power or kurtosis alone, while the adaptive constellations realize the predicted allocation gains. Moreover, target tracking alleviates the trade-off compared with detection.

\vspace{-0.1cm}
\section{System Model}
\vspace{-0.05cm}
\label{sec:system_model}
We consider a monostatic \ac{OFDM}-\ac{ISAC} system in which a single transmit signal is used for both \ac{SC}, i.e., to convey information to a communication receiver and to detect passive objects in a radar-like manner. The communication channel state information is assumed to be available at both the \ac{ISAC} transmitter and the communication receiver, e.g., through pilot-based channel estimation and feedback, enabling a channel-adaptive resource allocation at the transmitter. In contrast, the sensing channel is unknown and must be estimated by the sensing receiver from the received echo signal.

\subsection{\acs{ISAC} Transmitter}
The transmitter generates a single \ac{OFDM} symbol with $N$~subcarriers, indexed by \mbox{$n \in \{1,\ldots,N\}$}. Each subcarrier has an allocated transmit power $P_n$ and a constellation set \mbox{$\mathcal{X}_n = \{x_{n,1}, \ldots, x_{n,M}\}$} with \mbox{$|\mathcal{X}_n| \!=\! M$} symbols. The data symbol~$x_n$ is drawn as a realization of the random variable~$\RV{x}_n$ from $\mathcal{X}_n$ according to the \ac{PMF}~$\mathrm{Pr}_n(x_n)$ and is normalized such that \mbox{$\mathbb{E}\{\RV{x}_n\}\!=\!0$} and \mbox{$\mathbb{E}\{|\RV{x}_n|^2\}\!=\!1$}. The random transmit symbols~$\RV{s}_n$ are obtained by applying the subcarrier-dependent transmit power, i.e., \mbox{$\RV{s}_n\!=\!\sqrt{P_n}\RV{x}_n$}, with \mbox{$\sum_{n=1}^{N}\!P_n \!\leqslant\! P_{\text{tot}} \!=\! N$} and \mbox{$P_n\!\geqslant\!0$}. Finally, the \ac{OFDM} modulator transforms the symbols from the frequency domain into the time domain using an orthonormal \ac{IDFT} and prepends a \ac{CP} before the equivalent baseband signal is transmitted.

\subsection{Communication Channel, Receiver, and Metric}
For communications, we consider a frequency-selective block-fading channel, which is constant over one \ac{OFDM} symbol and impaired by \ac{AWGN}. After sampling, \ac{CP} removal, and an orthonormal \ac{DFT}, the received symbol on subcarrier~$n$ is \mbox{$\RV{y}^{\text{c}}_{n}=h^{\text{c}}_n\RV{s}_{n}+\RV{w}^{\text{c}}_{n}$}, where $h^{\text{c}}_n$ denotes the frequency-domain Rayleigh-fading channel coefficient, and \mbox{$\RV{w}^{\text{c}}_{n}\sim\mathcal{CN}(0,\sigma_{\text{c}}^2)$} is \ac{AWGN}. The communication performance on subcarrier~$n$ is assessed by the \ac{MI} conditioned on the known channel coefficient $h_n^{\text{c}}$ and the allocated power~$P_n$~\cite{Cover2006}
\vspace{-0.1cm}
\begin{equation} %
    \mathbb{I}(\RV{x}_n;\RV{y}^{\text{c}}_n \!\! \mid\! \!h_n^{\text{c}}, \!P_n) \!
    = \!
    \mathbb{E} \!
    \left\{\!
    \log_2 \!\!\left(\!\!
    \frac{
    f_{\RV{y}^{\text{c}}_n|\RV{x}_n,h_n^{\text{c}},P_n}
    (\RV{y}^{\text{c}}_n|\RV{x}_n,\!h_n^{\text{c}},\!P_n)
    }{
    f_{\RV{y}^{\text{c}}_n|h_n^{\text{c}},P_n}
    (\RV{y}^{\text{c}}_n|h_n^{\text{c}},\!P_n)
    }\!
    \right)
    \!\!\right\}\!\!,
    \label{eq:system_model:MI}
    \vspace{-0.1cm}
\end{equation}
which characterizes the asymptotically achievable rate with optimal coding. Here, $f_{\RV{y}^{\text{c}}_n|\RV{x}_n,h_n^{\text{c}},P_n}(\RV{y}^{\text{c}}_n|\RV{x}_n,h_n^{\text{c}},P_n)$ denotes the channel transition \ac{PDF}, and $f_{\RV{y}^{\text{c}}_n|h_n^{\text{c}},P_n}(\RV{y}^{\text{c}}_n|h_n^{\text{c}},P_n)$ denotes the \ac{PDF} of the received communication symbol on subcarrier~$n$.

\subsection{Sensing Channel, Receiver, and Metric}
For sensing, we assume a static channel consisting of $J$ point targets, indexed by \mbox{$j=1,\ldots,J$}, and a dominant \ac{SI} component, indexed by $j=0$, which dominates the received signal due to limited transmit-receive isolation in full-duplex \ac{ISAC} systems~\cite{wei_waveform_2026}. After sampling, \ac{CP} removal, and \ac{OFDM} demodulation, the received sensing symbol is
\mbox{$\RV{y}^{\text{s}}_{n}=h^{\text{s}}_{n}\RV{s}_{n}+\RV{w}^{\text{s}}_{n}$},
where \mbox{$\RV{w}^{\text{s}}_{n}\sim \mathcal{CN}(0,\sigma^2_{\text{s}})$} denotes \ac{AWGN}. The sensing channel is given by~\cite{braun_ofdm_2014}
\vspace{-0.15cm}
\begin{equation}
    h^{\text{s}}_{n} = \sum_{j=0}^{J} a_j \e^{\j\varphi_j}
    \e^{-\j 2\pi \frac{n}{N}\tilde{\tau}_j}, \quad \varphi_j \sim \mathcal{U}[0, 2\pi).
    \label{eq:system_model:sensing_channel_frequency_domain}
    \vspace{-0.15cm}
\end{equation}
Here, $a_j$ denotes the amplitude of target~$j$, with \mbox{$a_0 \gg a_j$}, \mbox{$j=1,\ldots,J$}, due to the dominant \ac{SI} component. The normalized delays $\tilde{\tau}_j$ are assumed to be integer multiples of the sampling interval and to lie within the \ac{CP} duration.

To estimate the sensing channel, we apply a sensing matched filter in the frequency domain
\vspace{-0.15cm}
 \begin{equation}
    \hat{\RV{h}}^{\text{s}}_{n}
    =\RV{y}^{\text{s}}_{n}\RV{s}_{n}^{\star} = h^{\text{s}}_{n} + h^{\text{s}}_{n} (P_n |\RV{x}_{n}|^2 - 1) + \RV{w}^{\text{s}}_{n} \sqrt{P_n}\RV{x}^\star_{n}.
    \label{eq:system_model:sensing_channel_estimate}
    \vspace{-0.15cm}
\end{equation} 

The channel estimate~$\hat{\RV{h}}^{\text{s}}[k]$ in the delay domain is obtained by applying an orthonormal \ac{IDFT} to the channel estimate~$\hat{\RV{h}}^{\text{s}}_{n}$ in the frequency domain. We quantify the sensing performance by the \ac{ISINR} of the \ac{TOI} with index $j_{\text{TOI}}$~\cite{geiger_constellation_2025}
\vspace{-0.15cm}
\begin{equation}
    \mathrm{ISINR}_{j_\text{TOI}} = \frac{|h^{\text{s}}[k_{j_\text{TOI}}]|^2 }{ \mathbb{E}\{ |\hat{\RV{h}}^{\text{s}}[k_{j_\text{TOI}}] - h^{\text{s}}[k_{j_\text{TOI}}]|^2 \} },
    \vspace{-0.15cm}
\end{equation}
where the expectation is taken with respect to the transmit symbols, sensing noise and random target phases. Here, $h^{\text{s}}[k_{j_\text{TOI}}]$ denotes the actual contribution of the \ac{TOI}. The \ac{ISINR} relates this contribution to the residual estimation error and thus serves as an indicator for detection performance~\cite{Geiger25JCS}.%

\subsection{Optimization Problem}
In this paper, we aim to jointly optimize the \ac{PA}, the subcarrier-dependent constellation sets, and the symbol \acp{PMF}, denoted by $\bm{P}$, $\boldsymbol{\mathcal{X}}$, and $\boldsymbol{\mathrm{Pr}}$, respectively, where the bold symbols indicate vectors. The objective is to maximize the total instantaneous \ac{MI} across all subcarriers while satisfying a total power constraint $P_{\text{tot}}$ and achieving a predefined sensing performance described by the \ac{ISINR} constraint $\Gamma$
\vspace{-0.15cm}
\begin{align}
    \max_{\bm{P},\,\boldsymbol{\mathcal{X}},\,\boldsymbol{\mathrm{Pr}}} \quad 
    & \frac{1}{N} \sum_{n=1}^{N}
    \mathbb{I}\!\left(\RV{x}_{n};\RV{y}^{\text{c}}_{n} \mid h_n^{\text{c}}, P_n\right),
    \label{eq:opt_full} \tag{P\num{1}} \\
    \text{s.t.} \quad 
    & \mathrm{ISINR}_{j_{\text{TOI}}}(\bm{P},\boldsymbol{\mathcal{X}},\boldsymbol{\mathrm{Pr}})\geqslant \Gamma, \label{eq:initial_C2}
      \tag{P\num{1}-C\num{1}} \\
    & \sum_{n=1}^{N} P_n \leqslant P_{\text{tot}}. %
      \tag{P\num{1}-C\num{2}}
\end{align}

\section{Decomposition into Constellation Shaping and Power-Kurtosis Allocation}
Directly solving~\eqref{eq:opt_full} is impractical, as it requires a joint optimization of the $N$~constellation sets, the $N$~symbol \acp{PMF}, and the \ac{PA}. To reduce the optimization complexity, we express both the sensing constraint and the communication objective in terms of the power and kurtosis per subcarrier. This enables a decomposition into an offline constellation-shaping and an online power-kurtosis allocation problem.

\subsection{Power-Kurtosis-Based Sensing Metric}
\label{sec:optimization_problem}
For arbitrary modulation and \ac{PA}, the \ac{ISINR} of the \ac{TOI} under dominant \ac{SI} can be approximated as~\cite{zhang_optimal_2025, meng_constellation_2026}
\vspace{-0.10cm}
\begin{equation}
    \mathrm{ISINR}_{j_{\text{TOI}}} \approx \frac{N a^2_{j_\text{TOI}}}{a^2_0 \cdot  \mathrm{SL}_{j_{\text{TOI}}} + \sigma_{\text{s}}^2},
    \vspace{-0.10cm}
\end{equation}
where the normalized \ac{SL} of the autocorrelation function is
\vspace{-0.10cm}
\begin{equation}
    \mathrm{SL}_{j_{\text{TOI}}} =
    \frac{1}{N}
    \underbrace{
    \left|
    \sum_{n=1}^{N}
    P_n \mathrm{e}^{-\mathrm{j}2\pi \frac{n}{N} \tilde{\tau}_{j_{\text{TOI}}}}
    \right|^2
    }_{\text{Deterministic}}
    +
    \frac{1}{N}
    \sum_{n=1}^{N} P_n^2(\kappa_n-1),
    \label{eq:optimization_sidelobe_lebel}
    \vspace{-0.15cm}
\end{equation}
with subcarrier-wise kurtosis \mbox{$\kappa_n\!=\!\mathbb{E}\{| \RV{x}_n |^4\}$}. A sensing optimal unit-modulus constellation and an \ac{MI}-maximizing circularly symmetric complex Gaussian correspond to \mbox{$\kappa\!=\!\num{1}$} and \mbox{$\kappa\!=\!\num{2}$}, respectively. Hence, we observe from~\eqref{eq:system_model:sensing_channel_estimate} and~\eqref{eq:optimization_sidelobe_lebel} that deviations from unit-modulus constellations, reflected by \mbox{$\kappa\!>\!1$}, as well as non-uniform \ac{PA}, introduce interference proportional to the sensing channel gain, which is dominated by the \ac{SI}.

If the delay of the \ac{TOI} is unknown, the local \ac{SL} is replaced by the \ac{AESL} over all non-zero delay bins~\cite{zhang_optimal_2025, meng_constellation_2026}
\vspace{-0.15cm}
\begin{equation}
    \mathrm{AESL} \!= \!\! \frac{1}{N\!\!-\!\!1} \! \! \left(\! \sum_{n=1}^{N} \!\!P_n^2 \!\! - \!\!\frac{1}{N}\! \left( \sum_{n=1}^{N} \!\!P_n\!\! \right)^{\!\!\!2}
    \right) \!\!+ \!\! \frac{1}{N} \!\! \! \sum_{n=1}^{N} \! \!P_n^2(\kappa_n\!\!-\!\!1)  .
    \label{eq:optimization:AESL}
    \vspace{-0.15cm}
\end{equation}

Previous work~\cite{zhang_optimal_2025,meng_constellation_2026,zhao_input_2026,yilmaz_rate_exponent_2026} has mainly considered the \ac{AESL}~\eqref{eq:optimization:AESL} as a sensing constraint, corresponding to an \textit{initial detection} mode for which a target should be detectable at \textit{any delay}. However, by averaging over the delay bins, the \ac{AESL}~\eqref{eq:optimization:AESL} discards the deterministic structure of the \ac{PA}-induced sidelobes in~\eqref{eq:optimization_sidelobe_lebel}, namely the \ac{DFT} of the \ac{PA}. We leverage this observation to formulate a \textit{tracking}-oriented sensing constraint, where prior delay information allows sensing to focus on a \textit{small delay neighborhood} $\mathcal{K}_{\text{tr}}$ around a predicted target position, with $|\mathcal{K}_{\text{tr}}|$ denoting the number of constrained delay bins. Hence, the optimal \ac{PA} for target detection is uniform~\cite{zhang_optimal_2025,meng_constellation_2026}, whereas tracking only requires suppressing the corresponding \ac{DFT} components of the \ac{PA}. This leaves additional degrees of freedom for communication-oriented \ac{PA}.

We note that both metrics depend only on the power~$P_n$ and kurtosis~$\kappa_n$ on each subcarrier, but not on the individual constellation points or their probabilities. Hence, the power- and constellation-dependent sensing constraint~\eqref{eq:initial_C2} can be reduced to the power-kurtosis constraint~\eqref{eq:optimization:modified_C1}, with the kurtosis restricted to \mbox{$1\leqslant\kappa_n\leqslant2$}, corresponding to a unit-modulus and a Gaussian modulation~\eqref{eq:optimization:modified_C3}.

\subsection{Power-Kurtosis-Based Communication Surrogate}
\label{subsec:optimization:upper_bound}
In contrast, the \ac{MI} objective in~\eqref{eq:opt_full} depends on the constellation points and their probabilities. To reduce the complexity of~\eqref{eq:opt_full}, we introduce a communication performance surrogate that, like the sensing metric, depends only on the power and the kurtosis per subcarrier. We obtain this surrogate by generalizing the upper bound from our previous work~\cite{geiger_constellation_2025} to include frequency-selective fading and \ac{PA}. This yields an allocation-oriented power-kurtosis formulation that complements the bound used for the input \ac{PDF} design in~\cite{zhao_input_2026}.

For a fixed communication noise variance, channel coefficient, and allocated power, maximizing the \ac{MI}
\mbox{$\mathbb{I}(\RV{x}_{n};\RV{y}^{\text{c}}_{n} \! \mid \! h_n^{\text{c}},\!P_n) \!= \! \mathbbm{h}(\RV{y}^{\text{c}}_{n} \! \mid \! h_n^{\text{c}},\!P_n) \! -\! \log_2(\pi \mathrm{e}\sigma_{\text{c}}^2)$}
is equivalent to maximizing the differential entropy~$\mathbbm{h}(\RV{y}^{\text{c}}_{n} \!\! \mid \!\! h_n^{\text{c}},\!P_n)$ of the received communication signal. Following~\cite{geiger_constellation_2025}, we map the power and kurtosis constraints on the transmit symbols to second- and fourth-order moment constraints on the received communication symbols
\vspace{-0.2cm}
\begin{align}
    C_{2,n} \!&= \!\mathbb{E}\!\left\{|\RV{y}^{\mathrm{c}}_{n}|^2    \right\} \!= \!|h^{\text{c}}_n|^2 P_n \!+ \!\sigma_{\mathrm{c}}^2,
    \label{eq:C1_channel} \\
    C_{4,n} \! &= \!\mathbb{E}\!\left\{         |\RV{y}^{\mathrm{c}}_{n}|^4 \right\} \!  = \!  \kappa_n |h^{\text{c}}_n|^4 P_n^2 \! \!+ \! 4 |h^{\text{c}}_n|^2 P_n  \sigma_{\mathrm{c}}^2 \!\! + \! 2\sigma_{\mathrm{c}}^4 .
    \label{eq:C2_channel}
    \vspace{-0.15cm}
\end{align}
Thus, we relax the \ac{MI} maximization to an entropy maximization problem with moment constraints $C_{2,n}$ and $C_{4,n}$, which depend only on the power~$P_n$ and the kurtosis~$\kappa_n$. The corresponding entropy-maximizing \ac{PDF} is~\cite{Cover2006}
\vspace{-0.1cm}
\begin{equation}
    f_{\RV{y}^{\text{c}}_{\text{max},n}}(y^{\text{c}}_n) = \exp\!\left(\gamma_{0,n} + \gamma_{2,n}|y^{\text{c}}_n|^2   + \gamma_{4,n}|y^{\text{c}}_n|^4 \right),
    \label{eq:entropy_maximizing_PDF}
\end{equation}
where the coefficients $\gamma_{0,n}$, $\gamma_{2,n}$, and $\gamma_{4,n}$ are chosen such that the \ac{PDF} is normalized and satisfies the moment constraints~\eqref{eq:C1_channel} and~\eqref{eq:C2_channel}. They can be obtained efficiently by one-dimensional root finding, as detailed in~\cite{geiger_constellation_2025}. Inserting \eqref{eq:entropy_maximizing_PDF} into the differential entropy yields the subcarrier-wise \ac{MI} upper bound
\vspace{-0.2cm}
\begin{equation}
    \hat{\mathbb{I}}_n(P_n,\kappa_n) \!=\! \frac{-\gamma_{0,n} \!\!-\! C_{2,n}\!\gamma_{2,n}\!\!-\! C_{4,n}\!\gamma_{4,n}}{\ln(2)} \! - \log_2\!\left(\pi \mathrm{e}\sigma_{\mathrm{c}}^2\right)\!.
    \label{eq:subcarrier_mi_bound}
    \vspace{-0.1cm}
\end{equation}
Note that this bound is not necessarily achievable, since the entropy-maximizing output \ac{PDF} is not guaranteed to be realizable under the considered system model~\cite{geiger_constellation_2025}.

\subsection{Optimization Problem Decomposition and Power-Kurtosis Allocation}
Both the sensing constraint and the communication surrogate depend only on the power and the kurtosis. We therefore decompose the original optimization problem~\eqref{eq:opt_full} into two parts: offline constellation shaping, see Sec.~\ref{sec:adaptive_constellation_shaping}, and online power-kurtosis allocation. During operation, the power and kurtosis on each subcarrier are selected by solving
\vspace{-0.15cm}
\begin{align}
    \max_{\bm{P},\,\bm{\kappa}} \quad
    & \frac{1}{N}\sum_{n=1}^{N}
    \hat{\mathbb{I}}_n(P_n,\kappa_n),
    \label{eq:opt_reduced} \tag{P\num{2}} \\
    \mathrm{s.t.} \quad
    & \mathrm{ISINR}_{j_{\text{TOI}}}(\bm{P},\bm{\kappa})\geqslant \Gamma,
      \tag{P\num{2}-C\num{1}} \label{eq:optimization:modified_C1}\\
    & \sum_{n=1}^{N} P_n \leqslant P_{\text{tot}}, %
      \tag{P\num{2}-C\num{2}} \\
    & 1\leqslant \kappa_n \leqslant 2,\quad \forall n ,
      \tag{P\num{2}-C\num{3}} \label{eq:optimization:modified_C3}
\end{align}
where the \ac{ISINR} constraint function~$\mathrm{ISINR}_{j_{\text{TOI}}}(\bm{P},\bm{\kappa})$ is chosen according to the sensing task: For detection, it is based on the \ac{AESL}~\eqref{eq:optimization:AESL}, whereas for tracking, it is based on the local \ac{SL}~\eqref{eq:optimization_sidelobe_lebel}. The problem contains only two scalar variables per subcarrier, $P_n$ and $\kappa_n$, and is optimized using sequential quadratic programming with multiple initializations. The optimized kurtosis~$\kappa_n$ selects the shaped constellation optimized offline, while $\sqrt{P_n}$ is applied as a scalar gain.

\section{Shaping Kurtosis-Adaptive Constellations\\for ISAC using an Autoencoder}
\label{sec:adaptive_constellation_shaping}

The online power-kurtosis allocation yields subcarrier-wise target kurtosis values~$\tilde{\kappa}_n$, which have to be realized by constellations. While previous work~\cite{Geiger25JCS,geiger_constellation_2025,du_reshaping_2023} optimizes individual constellations for fixed kurtosis constraints, this becomes impractical when constellations that can continuously adapt the kurtosis are required. We therefore extend the \ac{AE}-based constellation shaping framework from~\cite{geiger_constellation_2025} to optimize a kurtosis-conditioned family of constellations, \mbox{$\tilde{\kappa}_n \mapsto (\mathcal{X}_n(\tilde{\kappa}_n),\,\mathrm{Pr}_n(\cdot; \tilde{\kappa}_n))$}, from which constellations adapted to target kurtosis values can be generated. The general \ac{AE} setup and training procedure follow~\cite{geiger_constellation_2025} and the references therein. Here, we focus on the modifications required for kurtosis-adaptive constellation shaping. We replace the input-independent trainable parameters in~\cite{geiger_constellation_2025}, which describe the constellation points and probabilities, by \acp{NN} that take the target kurtosis $\tilde{\kappa}_n$ as input and output $M$ values, one for each constellation point.

Other work on adaptive constellations for varying \ac{SNR}~\cite{cammerer_trainable_2020} or phase noise~\cite{rode_end--end_2023} interprets the \ac{NN} outputs as Cartesian constellation coordinates. However, this approach fails to produce sensing optimal unit-modulus constellations, as shown in Sec.~\ref{sec:results}. We therefore parameterize the constellation set in polar form as \mbox{$\mathcal{X}_n(\tilde{\kappa}_n) \!=\! \left\{ r_{n,m}(\tilde{\kappa}_n) \exp\!\left(\mathrm{j}\theta_{n,m}(\tilde{\kappa}_n)\right) \right\}_{m=1}^{M}$}, and transform the \ac{NN} outputs in a kurtosis-dependent manner to obtain the radii, phases, and probabilities. The radii $\boldsymbol{r}_n$ are obtained from the radial \ac{NN} $f_{\mathrm{NN},\text{R}}(\tilde{\kappa}_n)$ as
\vspace{-0.15cm}
\begin{equation}
    \boldsymbol{r}_n(\tilde{\kappa}_n)
    \!=\!
    \max\!\left(
    \mathbf{1}\!+\!\sqrt{\tilde{\kappa}_n\!-\!1}\,
    \tanh\!\left(f_{\mathrm{NN},\text{R}}(\tilde{\kappa}_n)\right),
    \epsilon_{\text{R}}\mathbf{1}\!
    \right),
    \vspace{-0.1cm}
\end{equation}
where the maximum is applied element-wise. This yields unit-modulus constellations for \mbox{$\tilde{\kappa}_n=1$} and enables radial shaping for larger kurtosis values, while $\epsilon_{\text{R}}$ ensures a positive minimum radius. The phases $\boldsymbol{\theta}_n$ are obtained by adding the bounded output of the phase \ac{NN} $f_{\mathrm{NN},\theta}(\tilde{\kappa}_n)$ to a uniform angular grid
\vspace{-0.15cm}
\begin{equation}
    \boldsymbol{\theta}_n(\tilde{\kappa}_n) = \tilde{\boldsymbol{\theta}} +  \pi \cdot \tanh\!\left( f_{\mathrm{NN},\theta}(\tilde{\kappa}_n)\right),
    \vspace{-0.1cm}
\end{equation}
where \mbox{$[\tilde{\boldsymbol{\theta}}]_m=2\pi(m\!-\!1)/M$}. The probabilities are obtained by applying a softmax function to the scaled logits of the probability \ac{NN} $f_{\mathrm{NN},\text{Pr}}(\tilde{\kappa}_n)$
\vspace{-0.15cm}
\begin{equation}
    \mathrm{Pr}_n(\cdot;\tilde{\kappa}_n) \!=\! \mathrm{Softmax}\!\left(\min\!\left((\tilde{\kappa}_n\!\!-\!1)\!/\!\epsilon_{\text{Pr}},\!1\right) \!\cdot\! f_{\mathrm{NN},\text{Pr}}(\tilde{\kappa}_n) \right)\!,
    \vspace{-0.1cm}
\end{equation}
where \mbox{$0\!<\!\epsilon_{\text{Pr}}\!\ll \!1$} controls the transition to non-uniform probabilities. Thus, \mbox{$\tilde{\kappa}_n=1$} yields uniform probabilities, while larger target kurtosis values enable probabilistic shaping. Finally, the generated constellation is normalized to zero mean and unit average power, such that the allocated subcarrier power $P_n$ can be applied separately via a scalar gain.

\begin{figure}[!t]
    \centering
    \input{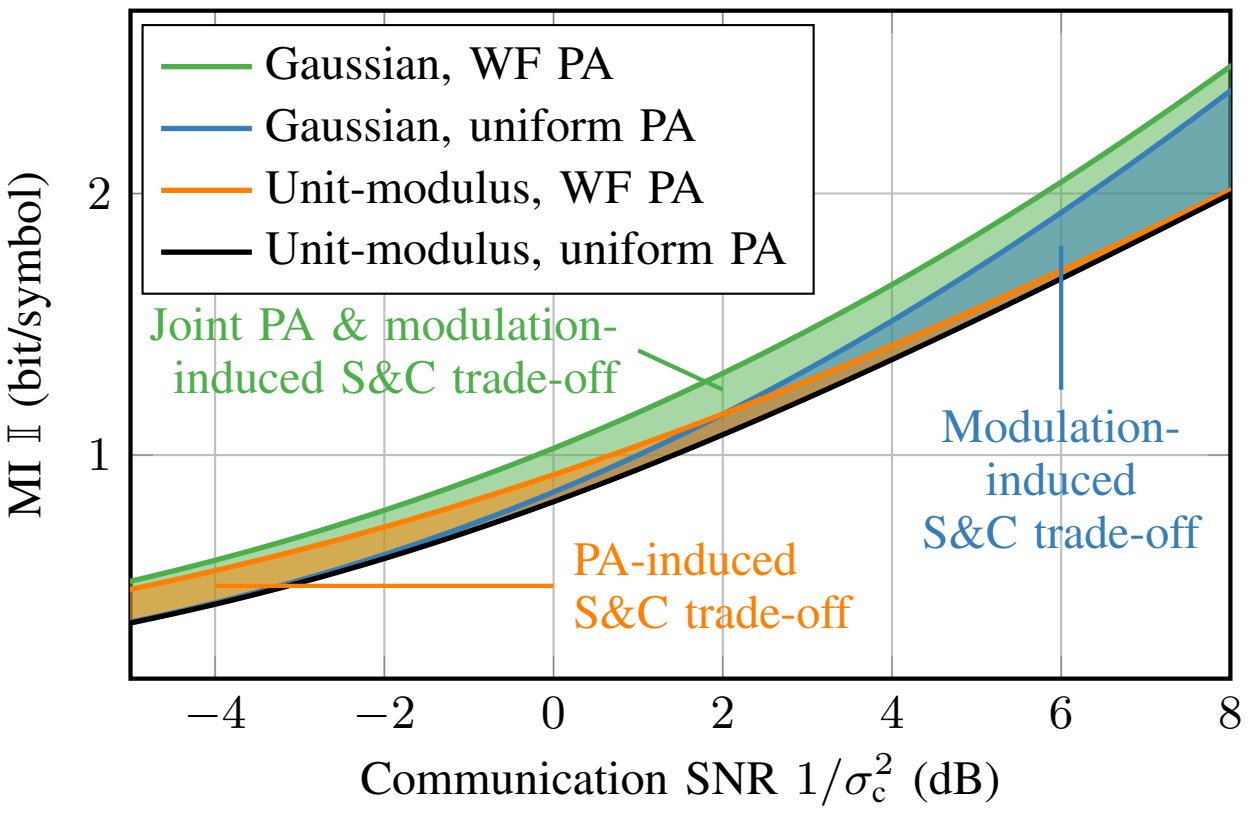}
    \vspace{-0.4cm}
    \caption{Impact of the four combinations of sensing- and communication-optimal modulation formats and \acp{PA} on the communication performance. The \ac{MI} of the Gaussian and unit-modulus reference inputs is obtained from the Shannon capacity and by numerically evaluating a \num{64}-\ac{PSK}, respectively.}
    \label{fig:results:trade_off_comparison}
    \vspace{-0.6cm}
\end{figure}

For training, we sample a batch of target kurtosis values and average the communication loss and kurtosis-constraint penalty from~\cite{geiger_constellation_2025} over the corresponding generated constellations. After training, the resulting constellation family can generate the constellation points and probabilities for the desired subcarrier-wise target kurtosis values $\tilde{\kappa}_n$ on the fly, avoiding the storage of separately optimized constellations.\footnote{The optimized kurtosis-adaptive constellation family is available at:\\ \url{https://github.com/kit-cel/isac_constellation_shaping}. (Upon acceptance)}

\begin{figure}[!t]
    \vspace{-0.22cm}
    \centering
    \input{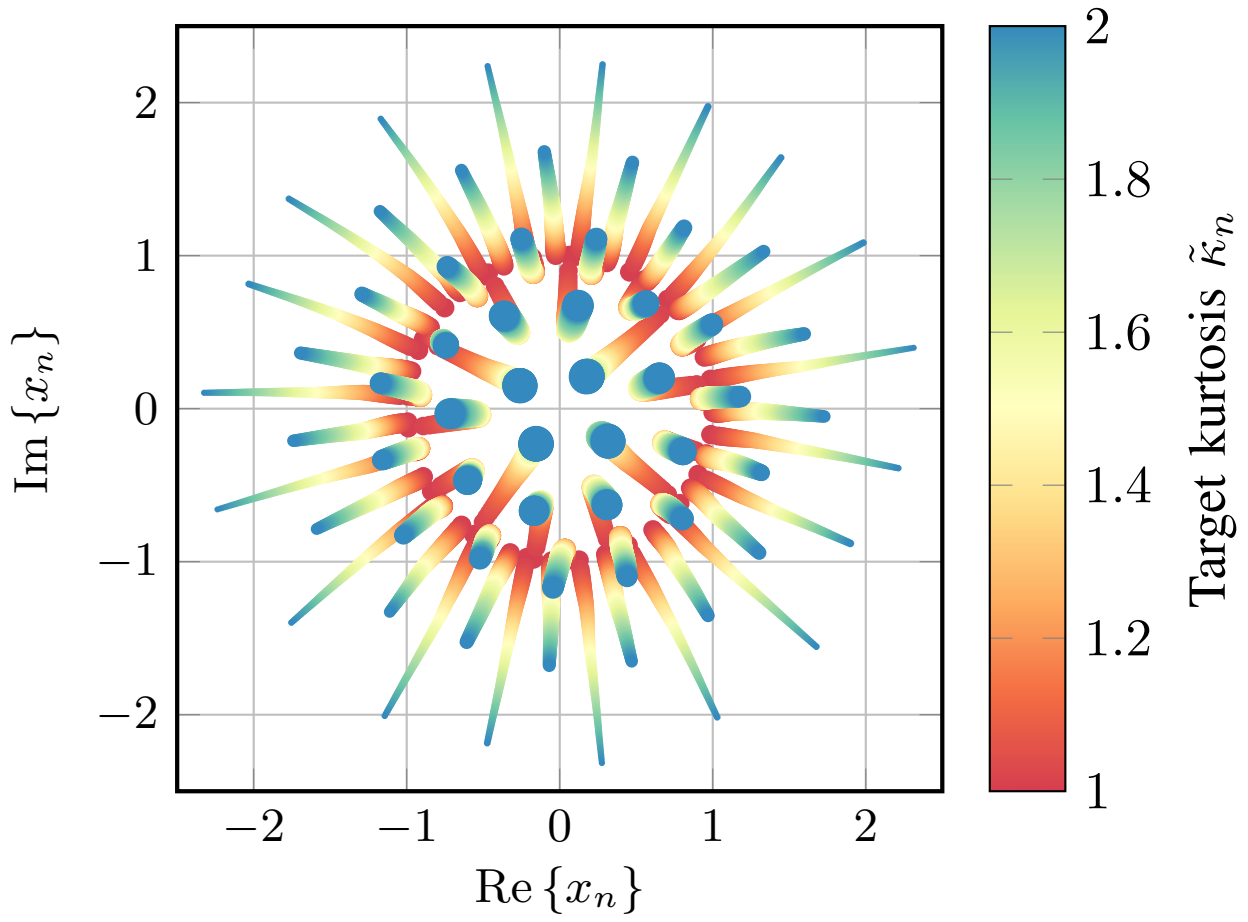}
    \vspace{-0.3cm}
    \caption{Optimized adaptive constellation with geometric and probabilistic shaping as a function of the target kurtosis~$\tilde{\kappa}_n$. The size of each constellation point indicates its probability.}
    \label{fig:adaptive_constellation}
    \vspace{-0.2cm}
\end{figure}

\section{Simulation Results}
\label{sec:results}
In this section, we evaluate the adaptive constellation, the power-kurtosis allocation, and their joint effect on the \ac{SC} trade-off. The \ac{OFDM} system uses \num{128} subcarriers, and the sensing channel consists of a dominant \ac{SI} component with \mbox{$a_0=\sqrt{10}$}, and a \ac{TOI} with \mbox{$a_1=1$}. The sensing and communication noise powers are \mbox{$\sigma^2_{\mathrm{s}}=\SI{3}{dB}$} and \mbox{$\sigma^2_{\mathrm{c}}=\SI{-2}{dB}$}.

All adaptive constellations contain \mbox{$M\!=\!\num{64}$} points and are shaped geometrically and probabilistically. The radius and phase \acp{NN} use five dense layers of sizes $(\num{1},\num{8},\num{16},\num{32},\num{64})$, while the probability \ac{NN} uses three dense layers of sizes $(\num{1},\num{8},\num{64})$, with ReLU activations in the hidden layers. We set $\epsilon_{\text{R}}\!=\!10^{-3}$ and \mbox{$\epsilon_{\text{Pr}}\!=\!0.01$}. Note that the \ac{NN} architectures are not optimized for minimal complexity, leaving room for further reduction and for studying the corresponding complexity-performance trade-off. During training, the communication \ac{SNR} is varied from \SI{0}{dB} to \SI{10}{dB}, and the batch size is increased from \num{5120} to \num{409600}, while the number of target kurtosis values is increased from \num{20} to \num{100} per batch. All results are averaged over \num{500} independent channel realizations.

\begin{figure*}[!b]
    \vspace{-0.45cm}
    \centering
    \setcounter{figure}{4}
    \input{figures/allocation}
    \vspace{-0.35cm}
    \caption{Jointly optimized power-kurtosis allocation for target detection and one channel realization. Subcarriers are sorted by decreasing communication channel gain~$|h_n^{\text{c}}|^2$. The heatmaps show allocated power~$P_n$ and kurtosis~$\kappa_n$ versus the \ac{ISINR} constraint~$\Gamma$.}
    \label{fig:results:power_kurtosis_allocation}
\end{figure*}

\subsection{Modulation- and Power-Induced \acs{ISAC} Trade-off}
We first analyze the modulation and \ac{PA}-induced \ac{SC} trade-off, shown in Fig.~\ref{fig:results:trade_off_comparison}. We observe that at low communication \ac{SNR}, the power-induced trade-off dominates and the modulation-induced trade-off vanishes, whereas at high \ac{SNR}, the opposite holds. Around the considered operating point of \SI{2}{dB} \ac{SNR}, both the \ac{PA} and modulation induce a pronounced \ac{SC} trade-off, motivating their joint optimization.

\begin{figure}[!t]
    \centering
    \setcounter{figure}{2}
    \input{figures/kurtosis_comparison}
    \vspace{-0.4cm}
    \caption{Achieved kurtosis~$\kappa_n$ versus target kurtosis~$\tilde{\kappa}_n$ for the proposed adaptive polar parameterization, the adaptive Cartesian parameterization, and kurtosis-specific constellations~\cite{Geiger25JCS}.}
    \label{fig:kurtosis_comparison}
    \vspace{-0.6cm}
\end{figure}

\subsection{Optimized Kurtosis-Adaptive Constellation Family}
Fig.~\ref{fig:adaptive_constellation} shows the geometrically and probabilistically shaped constellations generated by the optimized adaptive constellation family for different target kurtosis values~$\tilde{\kappa}_n$. We observe that the constellation smoothly transitions from a unit-modulus constellation at \mbox{$\tilde{\kappa}_n\!=\!\num{1}$} to a Gaussian-like distribution at \mbox{$\tilde{\kappa}_n\!=\!\num{2}$} by adapting both the constellation points and the \ac{PMF}.

Further, Fig.~\ref{fig:kurtosis_comparison} compares the achieved and target kurtosis for the proposed adaptive polar parameterization, the adaptive Cartesian parameterization, and kurtosis-specific constellations~\cite{Geiger25JCS}. The polar parameterization accurately generates constellations with the desired kurtosis over the full range. In contrast, the Cartesian parameterization fails to generate unit-modulus constellations, while the kurtosis-specific constellations partly undershoot the target kurtosis due to the sensitive balance between the \ac{SC} loss terms during optimization.

Next, Fig.~\ref{fig:comm_performance_verification} compares the communication performance of the proposed adaptive constellation with the upper bound derived in Sec.~\ref{subsec:optimization:upper_bound}, the kurtosis-specific constellations, and conventional modulation formats. The optimized adaptive constellation closely approaches the upper bound, with the remaining gap decreasing for lower \ac{SNR} values and increasing kurtosis. This confirms that the upper bound provides a suitable performance indicator for the power-kurtosis allocation. Moreover, the adaptive constellation achieves similar \ac{MI} to the kurtosis-specific constellations and outperforms conventional modulation formats. Thus, the adaptive constellation family offers the performance of kurtosis-specific optimizations while providing the flexibility required for the allocation.

\subsection{Kurtosis and Power Allocation}
Next, we analyze the jointly optimized power-kurtosis allocation for detection, shown in Fig.~\ref{fig:results:power_kurtosis_allocation}. When the sensing constraint is inactive (small $\Gamma$), we obtain the communication optimal \ac{WF} \ac{PA}, where the active subcarriers are assigned \mbox{$\kappa=2$}, corresponding to a Gaussian modulation. With stricter \ac{ISINR} constraints, the optimized allocation gradually becomes more sensing-centric. The \ac{PA} approaches a uniform distribution, while the kurtosis decreases towards \mbox{$\kappa=1$}, corresponding to unit-modulus constellations. Kurtosis outliers only occur on subcarriers with zero power, where the assigned kurtosis values are irrelevant. The optimizer first reduces~$\kappa_n$ on subcarriers with low channel gain~$|h^{\mathrm{c}}_n|^2$, which receive low power and therefore operate at low \ac{SNR}. In contrast, larger kurtosis values are retained on subcarriers with high channel gain and allocated power, which therefore operate at high \ac{SNR}, where increasing~$\kappa_n$ yields a larger \ac{MI} gain, as shown in Fig.~\ref{fig:comm_performance_verification}. We further observe a structured coupling between the allocated power $P_n$ and kurtosis $\kappa_n$, which could reduce the online optimization from two variables per subcarrier to one and thereby pave the way for low-complexity implementations.

\begin{figure}[!t]
    \vspace{-0.28cm}
    \centering
    \input{figures/Constellations_MI_comparison}
    \vspace{-0.8cm}
    \caption{Communication performance over an \ac{AWGN} channel of the novel adaptive constellations compared with the upper bound, kurtosis-specific optimized constellations, and the conventional modulation formats, \num{64}-\ac{PSK} and \num{64}-\ac{QAM}, for various communication \ac{SNR} values.}
    \label{fig:comm_performance_verification}
    \vspace{-0.45cm}
\end{figure}

\begin{figure}
    \centering
    \setcounter{figure}{5}
    \input{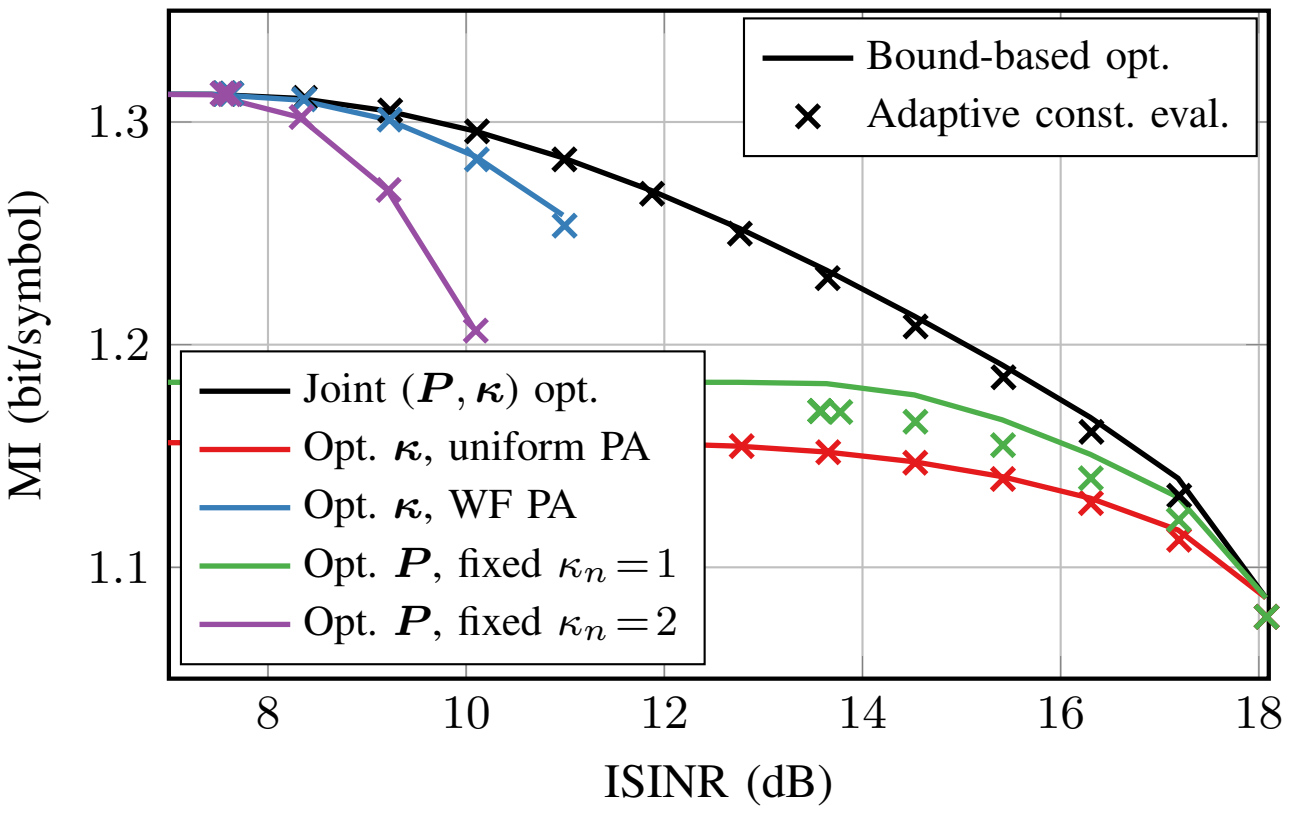}
    \vspace{-0.6cm}
    \caption{Detection-oriented \ac{SC} trade-off for joint and individual power-kurtosis optimization. Solid lines show the allocation results, i.e., the optimized upper bound~$\hat{\mathbb{I}}$ versus the imposed \ac{ISINR} constraint~$\Gamma$, while markers show the achieved \ac{MI}~$\mathbb{I}$ versus the achieved $\mathrm{ISINR}_{j_{\mathrm{TOI}}}$ using the optimized adaptive constellation family.}
    \label{fig:detection_trade_off}
    \vspace{-0.4cm}
\end{figure}

\subsection{\ac{ISAC} Trade-off for Detection and Tracking with Adaptive Constellations and Power-Kurtosis Allocation}

Finally, we compare in Fig.~\ref{fig:detection_trade_off} the detection-oriented \ac{SC} trade-off of the proposed joint power-constellation optimization with individual optimizations, where either the power or the kurtosis is optimized while the other is fixed. Solid lines denote the expected performance predicted by the bound-based power-kurtosis allocation, while markers show the achieved performance from the evaluation using the optimized adaptive constellation family. %

We observe that the individual optimizations with sensing-oriented designs, such as uniform \ac{PA} or unit-modulus constellations, are communication-limited, whereas communication-oriented designs, such as \ac{WF} \ac{PA} or Gaussian signaling, become infeasible under stricter sensing constraints. Notably, the joint optimization combines the advantages of both and achieves the best simultaneous \ac{SC} performance. Importantly, the markers largely follow the curves, with only a small gap at low kurtosis operating points, where the upper bound is known to overestimate the achievable \ac{MI}~\cite{geiger_constellation_2025}. This confirms that the proposed kurtosis-adaptive constellation family can realize the predicted power-kurtosis allocation gains and thereby bridges the gap between bound-based power-kurtosis allocation and practical finite constellation design.

\begin{figure}
    \centering
    \input{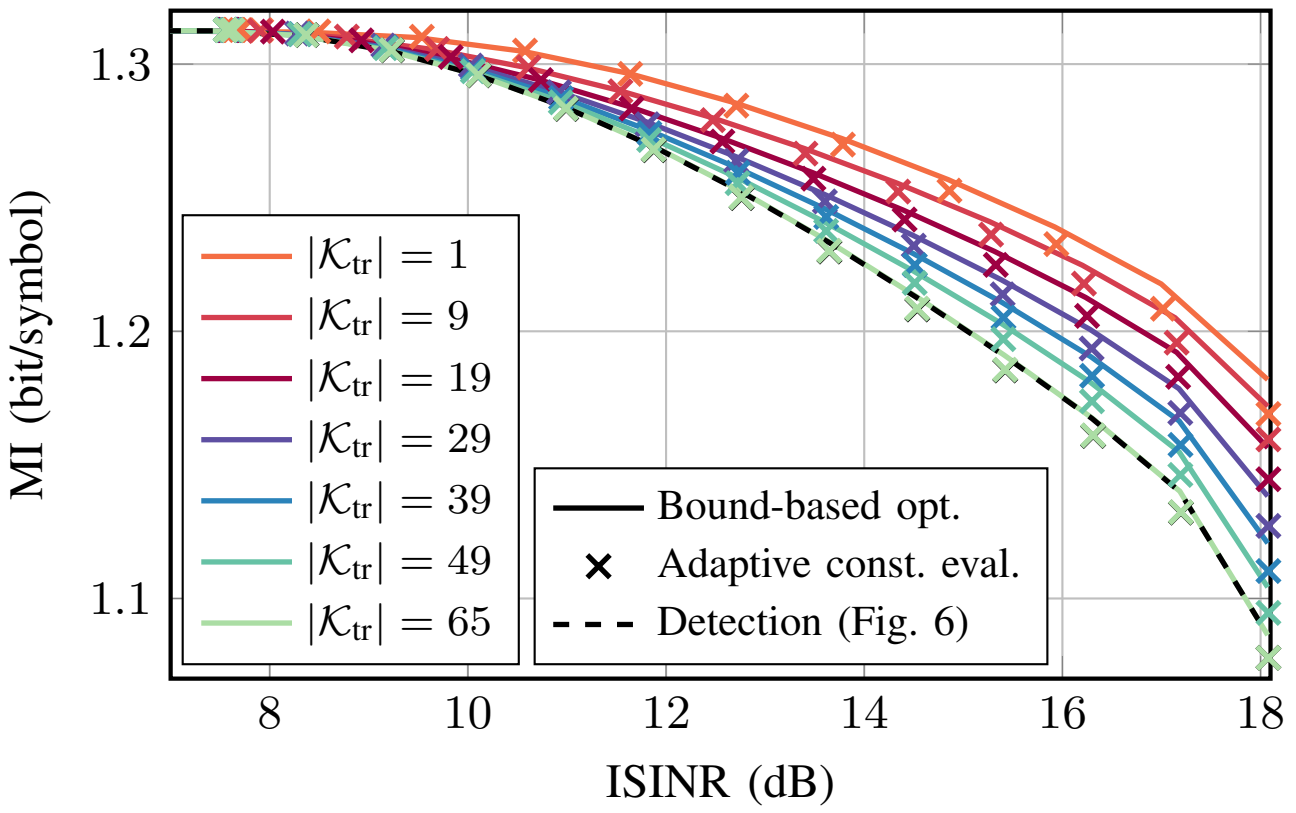}
    \vspace{-0.6cm}
    \caption{Tracking-oriented \ac{SC} trade-off for varying delay neighborhood sizes $|\mathcal{K}_{\text{tr}}|$ compared with detection. Solid lines show the allocation results, i.e., the optimized upper bound~$\hat{\mathbb{I}}$ versus the imposed \ac{ISINR} constraint $\Gamma$, while markers show the simulation results using the optimized adaptive constellation family, i.e., the achieved \ac{MI}~$\mathbb{I}$ versus the achieved~$\mathrm{ISINR}_{j_{\mathrm{TOI}}}$.}
    \label{fig:tracking_tradeoff}
    \vspace{-0.4cm}
\end{figure}

We finally investigate how prior target information changes the \ac{SC} trade-off by comparing detection and tracking in Fig.~\ref{fig:tracking_tradeoff}. Detection assumes no prior delay information and minimizes the delay-averaged \ac{SL}, whereas tracking assumes a prior target position and constrains only a local delay neighborhood $\mathcal{K}_{\mathrm{tr}}$ around it. As expected from the additional \ac{PA} degrees of freedom, tracking alleviates the \ac{SC} trade-off, especially for small $\mathcal{K}_{\text{tr}}$. For larger $|\mathcal{K}_{\text{tr}}|$, the performance approaches that of detection. Thus, prior target information should be exploited when available, with the sensing metric selected according to the sensing task.

\section{Conclusion}
In this work, we proposed a joint power-kurtosis allocation and adaptive-constellation framework for \ac{OFDM}-\ac{ISAC} systems. We showed that tracking alleviates the \ac{SC} trade-off compared with detection, highlighting the need for sensing task specific power-kurtosis allocation. The results further show that \ac{PA} and modulation dominate the \ac{SC} trade-off at low and high \ac{SNR}, respectively, and both must be optimized jointly at an intermediate \ac{SNR}. Here, the proposed joint allocation significantly outperforms an individual optimization of either power or modulation. Notably, the proposed polar parameterization yields an adaptive constellation family with the desired kurtosis and an \ac{MI} close to that of kurtosis-specific constellations. This makes the predicted allocation gains practically attainable and adaptive constellation shaping a promising candidate for future \ac{ISAC} systems that need to adapt dynamically to varying \ac{SC} requirements.

\end{document}